\documentclass[aip,amsmath,amssymb,reprint]{revtex4-1}

\usepackage{graphicx}
\usepackage{bm}

\usepackage{dcolumn}
\usepackage{inputenc}
\usepackage[T1]{fontenc}
\usepackage{mathptmx}
\usepackage{etoolbox}
\usepackage{xcolor}
\usepackage{tabularx}
\usepackage{enumitem}
\usepackage{float}
\usepackage[ruled,vlined]{algorithm2e}

\usepackage{subcaption} 
\usepackage{hyperref}
\usepackage{calc}
\usepackage{booktabs}
\usepackage{siunitx}
\usepackage{array}

\usepackage{amsfonts, amsthm, amssymb, amsmath}	
\usepackage{color}
\usepackage{graphicx}
\usepackage{mathrsfs}
\usepackage[english]{babel}
\usepackage{ upgreek }
\usepackage{ stmaryrd }
\usepackage{pgf,tikz}

\theoremstyle{plain}

\theoremstyle{remark}

\usepackage{tikz}
\usetikzlibrary{arrows.meta}
\usetikzlibrary{calc}

\begin{document}

\title[Low-rank approximation of Moment Tensor Potential enables reducing training set size without loss of accuracy]{Low-rank approximation of Moment Tensor Potential enables reducing training set size without loss of accuracy}

\author{Anna Bondarenko}
\affiliation{HSE University, Faculty of Computer Science, Pokrovsky boulevard 11, Moscow, 109028, Russian Federation}

\author{Nikita Rybin}
\affiliation{Moscow Engineering Physics Institute, Kashirskoe Highway 31, Moscow, 115409, Russian Federation}
\affiliation{Skolkovo Institute of Science and Technology, Skolkovo Innovation Center, Bolshoy boulevard 30, Moscow, 143026, Russian Federation}
\affiliation{Digital Materials LLC, Kutuzovskaya 4A, Odintsovo, 143001, Russian Federation}

\author{Maxim Rakhuba}
\affiliation{HSE University, Faculty of Computer Science, Pokrovsky boulevard 11, Moscow, 109028, Russian Federation}

\author{Ivan S. Novikov}
\email{isnovikov@hse.ru}
\affiliation{HSE University, Faculty of Computer Science, Pokrovsky boulevard 11, Moscow, 109028, Russian Federation}

\date{\today}

\begin{abstract}

In this study, we implement a low-rank approximation of Moment Tensor Potential (MTP) based on the tensor train (TT) decomposition. The implemented tensor-factorized MTP (TFMTP) and the original MTP model are actively trained via a MaxVol-based algorithm during molecular dynamics simulations of a four-component molten salt mixture, LiF-NaF-KF (FLiNaK), and geometry optimizations of a five-component Mo-Nb-Ta-W-V random alloy. We demonstrate that under a 1.5-fold compression, TFMTP requires two times fewer configurations for fitting than the original MTP model, while maintaining an indistinguishable level of accuracy. These actively trained MTP and TFMTP models are further used to evaluate the density and viscosity of FLiNaK at temperatures ranging from 600 to 1200 K, as well as the elastic constants and bulk modulus of the Mo-Nb-Ta-W-V alloy at zero temperature. For both atomic systems, the differences in physical properties predicted by MTP and TFMTP are negligible.

\end{abstract}

\maketitle

\section{Introduction}

Machine-learning interatomic potentials (MLIPs) \cite{behler2007-NNP,bartok2010-GAP,thompson2015-automated,shapeev2016-mtp,wang2018-deepmd,batzner2022-equivarnn,pun2019-pinn,drautz2019-ACE} have become a widely used tool in computational materials science \cite{jacobs2025_mlips}. These models use atomic positions, atomic types, and, optionally, lattice vectors of atomic structures as input parameters and predict energies of atomic configurations with high accuracy. As any machine-learning model, MLIPs have trainable parameters to be found during the optimization (or fitting) of MLIPs on the data typically calculated with density functional theory (DFT) calculations. MLIPs trained on DFT data enable predicting physical properties of materials for a reasonable time as they overcome the length and time-scale limitations of expensive DFT calculations.

In spite of impressive success in the application of MLIPs in computational materials science, they have drawbacks and one of them is the increase in the number of parameters as the number of atom types in the system grows or as the configuration space under investigation expands. The increasing number of parameters, in turn, leads to an increase in the size of the training set required to train MLIPs without overfitting which may be critical as DFT calculations used to generate training sets are computationally expensive. Therefore, it is necessary to develop methods for reducing MLIPs dimensionality without loss of accuracy.

One of the widely used methods to decrease the number of machine-learning model parameters is low-rank matrix or tensor approximation. This technique was successfully applied to reduce the complexity of neural networks and large language models \cite{lebedev2014speeding,novikov2015tensorizing,deng2020model,yin2021ttrec,hsu2022language,hu2022lora,li2023model,xiao2023comcat,wang2024svd,mo2025parameter}, e.g., TT-Rec \cite{yin2021ttrec} reduced the size of a deep-learning recommendation model by a factor of 2.6, corresponding to a 61.5\% reduction, while increasing validation accuracy from 80.45\% to 80.975\%, whereas COMCAT \cite{xiao2023comcat} reduced the number of parameters of DeiT-Base by 61.1\% while improving its ImageNet top-1 accuracy from 81.80\% to 82.26\%. In Ref. \cite{vorotnikov2025_low_rank_mtp}, two MLIPs, namely, Moment Tensor Potential (MTP) and Atomic Cluster Expansion (ACE), were compressed using low-rank matrix and tensor approximations, and it was demonstrated that the compressed models preserve the accuracy of the original (non-compressed) models when trained on the training set. However, in the mentioned paper, the question of the amount of data needed to train the original and compressed models was not investigated.

In this work, we apply an active learning algorithm proposed in Ref. \cite{podryabinkin2017_AL} to MTP and tensor-factorized MTP (TFMTP) introduced in Ref. \cite{vorotnikov2025_low_rank_mtp}, and investigate the number of configurations needed to fit MTP and TFMTP as well as the quality of the fitted models. The conceptual idea of low-rank compression of MTP is not new and was introduced in Ref. \cite{vorotnikov2025_low_rank_mtp} where every model was fitted to the same fixed dataset, and, therefore, it could only test accuracy at fixed data. Hence, the data question was left open. This study asks whether a compressed model needs less data when each model builds its own training set. The advances of the present work relative to Ref. \cite{vorotnikov2025_low_rank_mtp} are as follows. First, we explicitly implement the TFMTP model in the program code, whereas in Ref. \cite{vorotnikov2025_low_rank_mtp} TFMTP was not explicitly implemented. Second, we derive and implement the gradients of the loss function with respect to the radial parameters of TFMTP directly, whereas in Ref. \cite{vorotnikov2025_low_rank_mtp} these gradients were calculated via the gradients of the original MTP. This direct evaluation of parameter derivatives is what makes active learning applicable to TFMTP. The described developments enable us to apply the active learning algorithm to TFMTP, to construct training sets automatically for each type of MLIP, and to calculate physical properties of atomic systems with the actively trained models. As benchmarked systems, we consider the four-component molten salt mixture LiF-NaF-KF (FLiNaK) and the five-component Mo-Nb-Ta-W-V random alloy. We calculate the density and viscosity of FLiNaK at $T=600-1200$ K and the elastic constants and the bulk modulus of the near-equiatomic Mo$_{11}$Nb$_{11}$Ta$_{11}$W$_{11}$V$_{10}$ alloy at $T=0$ K. In our tests, we demonstrate that TFMTP, which has approximately 1.5 times fewer parameters than MTP, requires approximately two times fewer configurations for its fitting; however, its mentioned properties are close to the MTP ones. We also investigate the transferability of TFMTP and MTP. For the FLiNaK system, we test the transferability to a higher temperature of 1500 K. We demonstrate that both TFMTP and MTP fitted at $T=1200$ K are transferable to $T=1500$ K. For the Mo-Nb-Ta-W-V crystal, we consider eight additional compositions that were not present in the training set and test the transferability of TFMTP and MTP fitted only on Mo$_{11}$Nb$_{11}$Ta$_{11}$W$_{11}$V$_{10}$. These models extrapolate for all the unseen compositions. Therefore, we actively trained MLIPs on these new compositions to obtain reliable predictions. After active learning, both TFMTP and MTP gave bulk moduli close to each other and to the reference values obtained with the tabGAP model. Thus, our tests on unseen temperatures and compositions demonstrate that TFMTP preserves the predictive performance of MTP beyond the configurations emphasized during training. 

\section{Methodology}

\subsection{Moment tensor potential}

Here we describe Moment Tensor Potential (MTP) proposed in Ref. \cite{shapeev2016-mtp} for single-component materials and then generalized to the case of multi-component materials in Ref. \cite{gubaev2018-multMTP}. Let ${\bm x} = \{(\bm{l}_1, \bm{l}_2, \bm{l}_3); ({\bm r}_i,z_i), ~i=1,\ldots,N \}$ be a configuration with $N$ atoms in a supercell, each atom being encoded by its position ${\bm r}_i$, atomic type $z_i$. We denote the number of atomic types in a configuration by $N_T$. To approximate an infinite system, three lattice vectors $\bm{l}_1, \bm{l}_2, \bm{l}_3$ are also introduced for each configuration. Denote the energy of an atomic configuration $\bm x$ predicted with MTP by $E^{\rm MTP}$. This energy is the sum of contributions $V^{\rm MTP}(\mathfrak{\bm n}_i)$ of atomic neighborhoods ${\bf \mathfrak{n}}_i$ for $N$ atoms in a configuration:
\begin{align} \label{eq:EnergyMTP}
E^{\rm MTP} = \sum \limits_{i=1}^{N} V^{\rm MTP}(\mathfrak{\bm n}_i), 
\end{align}
i.e., MTP is a local potential. Each neighborhood is a tuple 
$$\mathfrak{ n}_i = ( \{{\bm r}_{i1},z_i,z_1 \}, \ldots, \{{\bm r}_{ij},z_i,z_j \}, \ldots, \{{\bm r}_{iN_ {\rm nbh}},z_i,z_{N_ {\rm nbh}} \} ),$$ 
where ${\bm r}_{ij} = {\bm r}_j - {\bm r}_i$ are relative atomic positions, $z_i$, $z_j$ are the types of central and neighboring atoms, and $N_ {\rm nbh}$ is the number of atoms in the neighborhood $\mathfrak{n}_i$ determined by a cutoff radius $R_{\rm cut}$, i.e., we include in the neighborhood each $j$-th atom with $|{\bm r}_{ij}| \leq R_{\rm cut}$. Each function $V^{\rm MTP}(\mathfrak{\bm n}_i)$ is a series of MTP basis functions $B_{\alpha}$:
\begin{align} \label{eq:SiteEnergyMTP}
V^{\rm MTP}({\bf \mathfrak{n}}_i) = \sum \limits_{\alpha} \xi_{\alpha} B_{\alpha}({\mathfrak{\bm n}}_i),
\end{align} 
where ${\bm \xi} = \{ \xi_{\alpha} \}$ are the linear parameters to be found. To construct the MTP basis functions we introduce the so-called moment tensor descriptors:
\begin{equation}\label{eq:MomentTesnsorDescriptors}
M_{\mu,\nu}({\mathfrak{\bm n}}_i)=\sum_{j=1}^{N_{\rm nbh}} f_{\mu}(|{\bm r}_{ij}|,z_i,z_j) r_{ij}^{\otimes \nu},
\end{equation}
where ``$\otimes$'' is the outer product of vectors and, therefore, ${\bm r}_{ij}^{\otimes \nu}$ is the tensor of the $\nu$-th order, which is the angular part describing many-body interactions, and $f_{\mu}(|{\bm r}_{ij}|,z_i,z_j)$ is the radial part describing only two-body interactions. The radial part has the following form:
\begin{align} \label{eq:RadialFunction}
\displaystyle
f_{\mu}(|{\bm r}_{ij}|,z_i,z_j) = \sum_{\beta=1}^{N_{\beta}} c^{(\beta)}_{\mu, z_i, z_j} T^{(\beta)} (|{\bm r}_{ij}|) (R_{\rm cut} - |{\bm r}_{ij}|)^2,
\end{align}
where $\mu$ is the number of the radial function $f_{\mu}$, ${\bm c}=\{c^{(\beta)}_{\mu, z_i, z_j}\}$ are the radial parameters to be found, $N_{\beta}$ is the number of Chebyshev polynomials $T^{(\beta)} (|{\bm r}_{ij}|)$.
 
We define the MTP basis function $B_{\alpha}$ as a contraction of one or more moment tensor descriptors \eqref{eq:MomentTesnsorDescriptors}, yielding a scalar, e.g.:
\[
\begin{aligned}
& B_0\left({\mathfrak{\bm n}}_{i}\right)=M_{1,0}\left({\mathfrak{\bm n}}_{i}\right), \\
& B_1\left({\mathfrak{\bm n}}_{i}\right)=M_{1,1}\left({\mathfrak{\bm n}}_{i}\right) \cdot M_{1,1}\left({\mathfrak{\bm n}}_{i}\right), \\
& B_2\left({\mathfrak{\bm n}}_{i}\right)=\left(M_{2,3}\left({\mathfrak{\bm n}}_{i}\right) M_{2,2}\left({\mathfrak{\bm n}}_i\right)\right) \cdot  M_{0,1}\left({\mathfrak{\bm n}}_{i}\right), \\
& \ldots
\end{aligned}
\]
However, the number of contractions yielding a scalar is infinite and, therefore, the number of MTP basis functions is infinite in general case. In order to restrict this number, we introduce the so-called level of the moment tensor descriptor:
\begin{equation} \label{eq:LevelMTD}
\displaystyle
{\rm lev} M_{\mu,\nu} = 2 + 4 \mu + \nu,
\end{equation}
and the level of the MTP basis function:
\begin{equation} \label{LevelMultMTD}
\displaystyle
{\rm lev} B_{\alpha} = {\rm lev} \underbrace {\prod_{p=1}^{P} M_{\mu_p,\nu_p}}_{\rm scalar} = \sum \limits_{p=1}^P (2 + 4 \mu_p + \nu_p).
\end{equation}
A set of MTP basis functions and, thus, a particular functional form of MTP depends on the maximum level, ${\rm lev_{\rm max}}$, which we also call the level of MTP. In the set of MTP basis functions, we include only those with ${\rm lev} B_{\alpha} \leq {\rm lev_{\rm max}}$. The level of MTP determines the number of linear parameters ${\bm \xi}$ and the number of radial functions $N_f$ (and, therefore, the number of the radial parameters $\bm c$) as the moment tensor descriptors depend on the number of the radial function $\mu$. For example, if we take ${\rm lev_{\rm max}}=8$ then we have nine basis functions $B_{\alpha}$ and $N_f=2$ radial functions (see Ref. \cite{novikov2020-mlip-2}).


\subsection{Tensor factorization of MTP}

A drawback of the functional of MTP is the tensor of the radial parameters $c^{(\beta)}_{\mu, z_i, z_j}$ as it has  $N_T^2 N_f N_b$ parameters and scales quadratically with the number of atomic types $N_T$ in a system. In Ref. \cite{vorotnikov2025_low_rank_mtp}, some MLIPs, including MTP, were compressed with low-rank matrix and tensor factorizations and it was demonstrated that it is possible to achieve up to 50\% compression without any loss of MTP accuracy. Among different compressed forms of MTP, tensor factorization of MTP (TFMTP) was introduced in  Ref. \cite{vorotnikov2025_low_rank_mtp}. TFMTP is based on tensor train (TT) decomposition (see, e.g., Ref. \cite{oseledets2011tensor}) of MTP radial parameters. In TFMTP, we express the tensor of the radial parameters as a contraction of four third-order core tensors $G^{(k)}$, $k=1,\ldots,4$:
\begin{equation} \label{eq:TT}
c^{(\beta)}_{z_i,z_j,\mu}
=
\sum_{\alpha_1=1}^{r_1}
\sum_{\alpha_2=1}^{r_2}
\sum_{\alpha_3=1}^{r_3}
G^{(1)}_{\,1,\,z_i,\,\alpha_1}
\;G^{(2)}_{\alpha_1,\,z_j,\,\alpha_2}
\;G^{(3)}_{\alpha_2,\,\mu,\,\alpha_3}
\;G^{(4)}_{\alpha_3,\,\beta,\,1}.
\end{equation}

The total number of the radial parameters in the TT format is $N_T r_1 + r_1 N_T r_2 + r_2 N_f r_3 + r_3 N_b$, i.e., if $r_1=r_2=r_3=r$ then we have $(N_T+N_f) r^2 + (N_T+N_b)r$ radial parameters instead of $N_T^2 N_f N_b$ in the original MTP. We emphasize that the only difference between MTP and TFMTP is the representation of the tensor of the radial parameters, and all the other steps for constructing MTP and TFMTP are similar.

We denote the parameters of MTP and TFMTP to be found by ${\bm \theta} = \{{\bm \xi}, {\bm c}\}$ and the energy of an atomic system predicted with these potentials by $E^{\rm MTP} = E^{\rm MTP}({\bm \theta})$ and $E^{\rm TFMTP} = E^{\rm TFMTP}({\bm \theta})$, respectively.

\subsection{Fitting}

Let $K$ be a number of configurations in a training set. To find optimal parameters ${\bar{{\bm {\theta}}}}$ of MTP and TFMTP, we minimize the following loss function:
\begin{equation} \label{eq:Loss}
\begin{array}{c}
\displaystyle
{\rm Loss}(\bm \theta)=\sum \limits_{k=1}^K \Bigl[ w_{\rm e} \left(E_k^{\rm MLIP}({\bm {\theta}}) - E^{\rm DFT}_k \right)^2 + 
\\
\displaystyle
w_{\rm f} \sum_{i=1}^{N} \sum_{l=1}^{3} \left( f^{\rm MLIP}_{i,l,k}({\bm {\theta}}) - f^{\rm DFT}_{i,l,k} \right)^2 +
\\
\displaystyle
w_{\rm s} \sum_{a=1}^6 \left( \sigma_{a,k}^{\rm MLIP}({\bm {\theta}}) - \sigma^{\rm DFT}_{a,k} \right)^2
\Bigr]  \to \operatorname{min},
\end{array}
\end{equation}
where $E^{\rm DFT}_k$, $f^{\rm DFT}_{i,l,k}$, and $\sigma^{\rm DFT}_{a,k}$ are the DFT energies, forces, and stresses to which we fit MLIP (MTP or TFMTP) energies $E_k^{\rm MLIP}({\bm {\theta}})$, forces $f_{i,l,k}^{\rm MLIP}({\bm {\theta}})$, and stresses $\sigma_{a,k}^{\rm MLIP}({\bm {\theta}})$, thus optimizing MLIP parameters ${\bm {\theta}}$. The factors $w_{\rm e}$, $w_{\rm f}$, and $w_{\rm s}$ in \eqref{eq:Loss} are non-negative weights which express the importance of energies, forces, and stresses with respect to each other. We find the optimal parameters ${\bar{{\bm {\theta}}}}$ numerically, using the BFGS algorithm. Before each fitting of MLIP, we start with a random initial guess.

We note that for the fixed radial parameters $\bm c$, the loss function \eqref{eq:Loss} is quadratic and convex in the linear MTP coefficients $\bm \xi$. 
However, joint optimization of $\bm \xi$ and $\bm c$ is generally nonconvex even for the original MTP. 
In TFMTP, the TT representation introduces additional
multiplicative dependencies between the core tensors
$G^{(k)}$, making the optimization problem nonconvex
with respect to these parameters as well. This additional nonconvexity appears to contribute to the observed sensitivity to scaling and initialization, which we address through the numerical adjustments described in the Computational details and Discussion sections.

\subsection{Active learning}

An algorithm for automated construction of a training set for MTP, namely, the active learning (AL) algorithm, was originally formulated for a single-component MTP in Ref. \cite{podryabinkin2017_AL} and then generalized to the case of many components in Ref. \cite{gubaev2018-multMTP}. Since the difference between MTP and TFMTP is only in the representation of the radial parameters, the algorithm proposed in Ref. \cite{gubaev2018-multMTP} can be also applied to TFMTP and, therefore, we describe it in general form, using the MLIP abbreviation.

Assume we have an initial training set including $K$ configurations and we found the vector of $m$ optimal MLIP parameters ${\bm \bar{\theta}}$ after solving \eqref{eq:Loss}. We then construct a matrix

\[
B=\left(\begin{matrix}
\frac{\partial E^{\rm MLIP}_1}{\partial \theta_1}({\bm {\bar{\theta}}}) & \ldots & \frac{\partial E^{\rm MLIP}_1}{\partial \theta_m}({\bm {\bar{\theta}}}) \\
\vdots & & \vdots \\
\frac{\partial E^{\rm MLIP}_K}{\partial \theta_1}({\bm {\bar{\theta}}}) & \ldots & \frac{\partial E^{\rm MLIP}_K}{\partial \theta_m}({\bm {\bar{\theta}}}) 
\end{matrix}\right).
\]
Next, we find a submatrix $A$ of size $m \times m$ from the matrix $B$ with the maximum absolute value of the determinant (maximum volume) $|{\rm det}(A)|$, i.e., we select a set of the $m$ most linearly independent rows. To find a submatrix $A$ of maximum volume, we apply the MaxVol algorithm~\cite{goreinov2010_maxvol}. This approach guarantees that the selected rows of $A$ correspond to geometrically different configurations, which makes it possible to cover a part of the configuration space of interest.

After training the initial MLIP and constructing the matrix $A$, we start any atomistic simulation and compute the extrapolation grade for each configuration ${\bm x^*}$ created during this simulation
\begin{equation}
\label{eq:extrapol_grade}
    \gamma({\bm x^*}) = \max\limits_{1 \le j \le m} |c_j|,
\end{equation}
where the vector $\textbf{c} = (c_1, \ldots, c_m)$ for the configuration ${\bm x^*}$ is
\begin{equation}
\label{eq:c}
    \textbf{c} = \Bigl(\frac{\partial E^{\rm MTP}}{\partial \theta_1} (\bm {\bar{\theta}}, {\bm x}^*), \dots, \frac{\partial E^{\rm MTP}}{\partial \theta_m} (\bm {\bar{\theta}}, {\bm x}^*) \Bigr) A^{-1}.
\end{equation}
The extrapolation grade $\gamma$ quantifies the increase in $|{\rm det}(A)|$ when a candidate configuration ${\bm x^*}$ is added to the training set. During a simulation, we create a pool of preselected candidate configurations (or, the preselected set) that may be added to the training set. To that end, we define the lower bound, $\gamma_{\rm low}$, and the upper bound, $\gamma_{\rm up}$, of permissible extrapolation. If $\gamma_{\rm low} \leq \gamma({\bm x^*}) \leq \gamma_{\rm up}$ then we add the configuration ${\bm x^*}$ to the preselected set and continue an atomistic simulation, as the errors in predicting energies, forces, and stresses by MLIP are still acceptable due to permissible extrapolation. If $\gamma({\bm x^*}) > \gamma_{\rm up}$ then the extrapolation is too risky and, therefore, we terminate an atomistic simulation, add this configuration to the preselected set, and apply the MaxVol algorithm to update the matrix $A$ by selecting the most diverse configurations from the preselected set. In other words, we select the new configurations to be added to the training set. These selected configurations are then evaluated via density functional theory (DFT) calculations, added to the training set, and used to re-fit MLIP. We then restart the entire procedure -- a simulation with extrapolation control, selection, DFT calculations, re-training -- and repeat it iteratively until no configurations are preselected during atomistic simulation.

An important hyperparameter of the formulated AL algorithm is the size $m \times m$ of the matrix $A$ where $m$ is the number of parameters in MLIP. As shown in Figure 8 of Ref.~\cite{novikov2018_RPMD_AL_MTP}, the number of configurations selected by the aforementioned AL algorithm for fitting of MTP grows with the number of model parameters. However, in the mentioned study, the parameter growth stemmed from raising the potential level, i.e., adopting a more complex functional form, which concurrently enhanced the accuracy of the potential. In this study, we investigate the number of configurations required to train MTP and TFMTP of the same level and demonstrate that the TFMTP model requires less data than MTP but is nearly as accurate as the original MTP model.

\section{Results and discussion}

\subsection{Computational details}

We tested the amount of data needed to actively train MTP and TFMTP using two systems: the four-component eutectic molten salt LiF-NaF-KF (46.5-11.5-42 mol\%) mixture and the five-component Mo-Nb-Ta-W-V alloy. Active learning was carried out during two widely-used atomistic simulations: molecular dynamics (MD) for LiF-NaF-KF (FLiNaK) and geometry optimization for Mo-Nb-Ta-W-V structures. Selected configurations were calculated within the DFT framework (with Perdew-Burke-Ernzerhof exchange-correlation functional~\cite{PhysRevLett.77.3865}) using the Quantum Espresso package \cite{giannozzi2009_qe}. We followed the recommendations of the standard solid-state pseudopotential project~\cite{prandini2018precision} and used pseudopotentials from GBRV~\cite{garrity2014pseudopotentials} and Pslibrary~\cite{dal2014pseudopotentials} for modeling FLiNaK. Mo-Nb-Ta-W-V was modeled using projector-augmented-wave~\cite{blochl1994projector} pseudopotentials from the PseudoDojo library~\cite{van2018pseudodojo}. In the case of FLiNaK, we also applied DFT-D3 correction with zero-damping function~\cite{grimme2010consistent}, which is vital for obtaining the density of molten FLiNaK in reasonable agreement with experimental data~\cite{rybin2024-FLiNaK}. Other details of the DFT calculations are given in Table \ref{tab:dataset_generation}.

\begin{table}[!h]
    \caption{Computational details on the generation of the training sets.}
    \begin{center}
    \begin{tabular}{c|c|c|c|c}
    \hline
    \hline
    System & Level & Cutoff, & k-mesh & Cell size,\\
     & of theory & Ry & size &  \#atoms
    \\
    \hline
    FLiNaK & PBE+D3 & 70 & 2 $\times$ 2 $\times$ 2 & 56 
    \\
    MoNbTaWV & PBE & 50 & 4 $\times$ 4 $\times$ 4 & 54 
    \\
    \hline
    \hline
    \end{tabular}
    \end{center}
    \label{tab:dataset_generation}
\end{table}

We investigated the FLiNaK system in two stages. At the first stage, we used the MatterSim model (see Ref. \cite{yang2024mattersim}) instead of DFT to calculate energies, forces, and stresses of the selected configurations to test our assumption that TFMTP requires less data than MTP. To that end, we took MTP and TFMTP of the 14-th level (i.e,, $N_f = 4$ and 52 linear parameters) with $N_{\beta} = 8$ Chebyshev polynomials and $R_{\rm cut} = 5$ \AA. The ranks for TFMTP were $(r_1,r_2,r_3)=(4,8,4)$ and, thus, MTP of 14-th level has 568 parameters, while TFMTP of the same level has 360 parameters, i.e. approximately 1.5 times less. For constructing an initial training set, we conducted 1 ps of MD simulations with a time step of 1 fs in the NVT ensemble at $T=600$ K with a supercell including 56 atoms and a size of 9.63 \AA. We utilized the MatterSim model and sampled a trajectory including 1000 configurations. We then took each 50-th configuration from the trajectory and formed the initial training set. Next, after fitting of MTP and TFMTP on this initial dataset, we selected additional configurations from the mentioned trajectory with the MaxVol algorithm, added them to the initial training set, obtained training sets for both of the models, and re-trained these models. Finally, starting from the obtained training sets, we actively trained ensembles of five MTP and TFMTP during MD simulations of 100 ps in the NVT ensemble at $T=600$ K and compared the number of selected configurations. We also compare average values and standard deviations of the loss function \eqref{eq:Loss} as well as energy, force, and stress root-mean-square errors calculated on the training sets selected with MTPs. At the second stage, we used MTP and TFMTP with the same hyperparameters instead of the level -- here we took the 16-th level and, therefore, TFMTP and MTP have 400 and 608 parameters, respectively. At this stage, we created the training sets with DFT calculations using parameters provided in Table \ref{tab:dataset_generation}. An initial MD trajectory including 1000 configurations was generated during 1 ps of MD simulations in the NVT ensemble at $T=1200$ K and, as at the first stage, each 50-th configuration from this trajectory was added in the initial training set. After that, additional configurations were selected from the MD trajectory with the MaxVol algorithm using both MTP and TFMTP and added in the training sets for fitting of these models. Finally, starting from the created datasets, we actively trained MTP and TFMTP during MD simulations of 150 ps in the NPT ensemble at $T=1200$ K and obtained the resulting training sets for MTP and TFMTP. We also tested the transferability of the fitted MTP and TFMTP models to a higher temperature of 1500 K. As opposed to the first stage, here we actively trained only one model of each type due to the computational expense of DFT calculations in comparison with MatterSim. However, we additionally passively trained four more models of the same types on the obtained training sets. We then compared the number of configurations in the training sets obtained with actively trained models and average values of the loss function \eqref{eq:Loss} on the training set created with MTP, average density, and average viscosity of FLiNaK at $T=600-1200$ K predicted with the ensembles of MLIPs. To compute the density, we conducted MD simulations of 120 ps in the NPT ensemble for a supercell of 1512 atoms with each of the fitted potentials, and then averaged the density over the ensemble of MLIPs. To calculate viscosity, we utilized the Green-Kubo (GK) approach \cite{green1954markoff} in which viscosity is calculated through the integral of the auto-correlation function of the off-diagonal elements of the pressure tensor:

\begin{equation} \label{eq:viscosity}
    \eta = \dfrac{V}{k_B T} \int \limits_{0}^{\infty} <P_{\alpha \beta} (t) P_{\alpha \beta} (0)> dt,
\end{equation}
where $\eta$ is the viscosity, $k_B$ is the Boltzmann constant, and $P_{\alpha \beta}$ are the off-diagonal elements of the pressure tensor. Here, we ran 10 trajectories for each of five MLIPs in the ensemble starting from different velocities. To sample the configurations for the viscosity calculation, we first equilibrated a configuration of 3584 atoms for 200 ps in the NVT ensemble and then ran 1.5 ns of simulations in the NVE ensemble with a 50 ps auto-correlation time to ensure the convergence of viscosity. The resulting viscosity was averaged over 50 trajectories for each type of the model.

In addition to comparing MTP and TFMTP with each other, we also compared their average density and viscosity with the experimental ones. Furthermore, we calculated density using one of the foundation models —- the MACE model —- and demonstrated that the broadly pretrained MACE-MP-0 model \cite{batatia2025_mace_mp_0}, which includes 3847696 trainable parameters and has a cutoff radius of 6 \AA, yielded density less accurate with respect to experiment than MTP/TFMTP did. At the same time, MACE-MP-0 fine-tuned on the training set automatically constructed during the active learning of TFMTP predicted density similar to that of MTP/TFMTP. Finally, we also analyzed the performance of the TFMTP, MTP, and MACE models.

We started investigating the Mo-Nb-Ta-W-V from the near-equiatomic Mo$_{11}$Nb$_{11}$Ta$_{11}$W$_{11}$V$_{10}$ crystal. We tested active learning of TFMTP and MTP during geometry optimization (relaxation) of this random alloy. In this test, we utilized MTP of the 20-th level (i.e., $N_f = 5$ and 288 linear parameters) with $N_{\beta} = 8$ Chebyshev polynomials, and $N_T=5$ and, thus, MTP has 1293 parameters. For TFMTP we took $(r_1,r_2,r_3)=(5,10,5)$ and it has 858 parameters, approximately 1.5 times less than in MTP, as in the test with the FLiNaK system. To create an initial training set, we utilized a near-equiatomic bcc supercell with the size of 9.585 \AA ~including 54 atoms, namely, 11 atoms of Mo, Nb, Ta, and W, and 10 atoms of V. We generated 10 configurations with random arrangements of Mo, Nb, Ta, W, and V atoms at the supercell nodes and prerelaxed them with DFT using the $2 \times 2 \times 2$ k-mesh. Next, we perturbed the atomic coordinates 10 times for each prerelaxed configuration by applying small random displacements drawn from a Gaussian distribution with a standard deviation of 0.05 \AA ~and calculated an initial training set including 110 configurations with DFT using the parameters from Table \ref{tab:dataset_generation}. Finally, we actively trained MTP and TFMTP during the geometry optimization of these 110 structures starting from the created initial training set. We compared the sizes of the training sets created during the active learning of MTP and TFMTP, as well as the values of the loss function \eqref{eq:Loss} and the energy, force, and stress errors calculated on the MTP training set, elastic constants, and bulk moduli calculated with the fitted models.

After active training of MTP and TFMTP on the near-equiatomic Mo$_{11}$Nb$_{11}$Ta$_{11}$W$_{11}$V$_{10}$ crystal, we tested their transferability to unseen compositions of Mo-Nb-Ta-W-V. For this purpose, we took eight following compositions of Mo-Nb-Ta-W-V: $$
\begin{aligned}
&\text{Mo}_{15}\text{Nb}_{10}\text{Ta}_{9}\text{W}_{18}\text{V}_{2}, \quad \text{Mo}_{17}\text{Nb}_{15}\text{Ta}_{1}\text{W}_{15}\text{V}_{6}, \\
&\text{Mo}_{15}\text{Nb}_{14}\text{Ta}_{10}\text{W}_{12}\text{V}_{3}, \quad \text{Mo}_{24}\text{Nb}_{7}\text{Ta}_{8}\text{W}_{2}\text{V}_{13}, \\
&\text{Mo}_{2}\text{Nb}_{2}\text{Ta}_{26}\text{W}_{10}\text{V}_{14}, \quad \text{Mo}_{15}\text{Nb}_{12}\text{Ta}_{15}\text{W}_{2}\text{V}_{10}, \\
&\text{Mo}_{3}\text{Nb}_{6}\text{Ta}_{26}\text{W}_{2}\text{V}_{17},  \quad \text{Mo}_{21}\text{Nb}_{16}\text{Ta}_{11}\text{W}_{4}\text{V}_{2}.
\end{aligned}
$$ These compositions are shown in Fig. 1 from Ref. \cite{byggmastar2021_monbtawv}. As for Mo$_{11}$Nb$_{11}$Ta$_{11}$W$_{11}$V$_{10}$, we generated 10 configurations with random arrangements of Mo, Nb, Ta, W, and V atoms for each of the mentioned compositions and prerelaxed them with DFT using the $2 \times 2 \times 2$ k-mesh.  Next, we estimated the extrapolation grades of the prerelaxed structures and found that both MTP and TFMTP extrapolate for all the unseen compositions of Mo-Nb-Ta-W-V. For this reason, we actively trained our models during relaxation of these eight compositions and showed that the number of configurations in the resulting training sets still grows proportionally to the number of parameters in TFMTP and MTP. Moreover, we calculated bulk moduli with the actively trained models and compared them with the ones provided in Ref. \cite{byggmastar2021_monbtawv} and computed with the tabGAP model.

Finally, we describe some details of the potentials training. In the case of fitting on the MatterSim-labeled data, we took the standard for MTP weights: $w_{\rm e}=1$ eV$^{-2}$, $w_{\rm f}=0.01$ (eV/\AA)$^{-2}$, and $w_{\rm s}=0.001$ eV$^{-2}$ in the loss function \eqref{eq:Loss}. In the case of fitting on DFT data, we took the weights $w_{\rm e}=0.1$ eV$^{-2}$, $w_{\rm f}=0.001$ (eV/\AA)$^{-2}$, and $w_{\rm s}=0.0001$ eV$^{-2}$ for the FLiNaK system, and $w_{\rm e}=0.1$ eV$^{-2}$, $w_{\rm f}=0.001$ (eV/\AA)$^{-2}$, and $w_{\rm s}=0.001$ eV$^{-2}$ for the Mo-Nb-Ta-W-V crystal. Furthermore, to fit MLIPs on the FliNaK system, we subtracted the minimum energy of configurations in the initial training sets and fitted to the subtracted energies due to an unresolved problem with the scaling of TFMTP (see the Discussion section). However, it is suitable only for FLiNaK as this system includes only one composition. For the Mo-Nb-Ta-W-V crystal including different compositions, we calculated the energies of Mo, Nb, Ta, W, and V in vacuum with DFT and subtracted the energy of each isolated atom from the total energy of a configuration, i.e. we fitted to $\tilde{E}^{\rm DFT} = E^{\rm DFT} - \sum_{i=1}^5 N_{z_i} E^{\rm DFT}_{{\rm vac}, z_i}$, where $N_{z_i}$ is the number of atoms of type $z_i$ and $E^{\rm DFT}_{{\rm vac}, z_i}$ is the DFT energy of an isolated atom of type $z_i$. The number of BFGS iterations was limited to 10000, and the thresholds for active learning were $\gamma_{\rm low} = 2$ and $\gamma_{\rm up} = 10$. 

\subsection{Results for FLiNaK}

\subsubsection{Training set size and accuracy on MatterSim data}

To understand the prospects of active learning of the compressed potentials, namely, TFMTP, for reducing the number of configurations in a training set, we actively trained ensembles of five MTPs and TFMTPs during MD simulations of the FLiNaK system in the NVT ensemble at $T = 600$ K and used the foundation MatterSim model to calculate training sets. The results are given in Table \ref{tab:dataset_size_on_MatterSim}. TFMTPs have 1.5 times fewer parameters than MTPs, but selected more than 1.5 times fewer configurations than MTPs which confirms our main assumption: since the number of parameters in MLIP is a hyperparameter of the MaxVol algorithm used to automatically create a training set, the number of selected configurations reduces as this hyperparameter decreases. We also note that the standard deviations from the average numbers of the selected configurations are quite small.

\begin{table}[!ht]
\caption{Number of parameters and average number of configurations in the training sets for the ensembles of the 14-th level MTP and TFMTP actively trained on the data calculated with MatterSim. The number of selected configurations is given with 2-$\sigma$ confidence interval. The ensemble of TFMTPs has 1.5 times less parameters and requires more than 1.5 times fewer configurations for its fitting than the ensemble of MTPs.}
\label{tab:dataset_size_on_MatterSim}
\centering
\begin{tabular}{ccccc}
\hline \hline
Model & \# parameters & \# selected configurations \\
\hline 
TFMTP & 360 & $927 \pm 20$ \\

MTP & 568 & $1608 \pm 36$ \\
\hline \hline
\end{tabular}
\end{table}

We then examined the accuracy of the fitted models. To that end, we took five training sets automatically created with active learning of five MTPs and then estimated the average values of the loss function \eqref{eq:Loss} and the average root-mean-square errors (RMSEs) for energies, forces, and stresses. Namely, for each of the five TFMTPs we calculated the above values on the training sets created with the five MTPs and then calculated the average values and their standard deviations. We did the same computations with MTPs. The results are given in Table \ref{tab:accuracy_on_MatterSim}. We observe that the values of the loss functions and RMSEs are close to each other. Thus, Tables \ref{tab:dataset_size_on_MatterSim} and \ref{tab:accuracy_on_MatterSim} indicate that the TT-based factorization eliminates parametric redundancy without any significant loss in the potential's intrinsic accuracy. Therefore, a reasonable compression of the MTP model corresponding to the reduction of about 50 \% of the radial parameters or smaller (see Ref. \cite{vorotnikov2025_low_rank_mtp}) leads to a reduced training set without a loss of the model's accuracy. From this test, we conclude that TFMTP is indeed a prospective model in terms of reducing the training set without a loss of accuracy, and we further compare MTP and TFMTP actively trained on DFT calculations.

\begin{table}[!ht]
\caption{Average values of the loss function and root-mean-square errors for energies, forces, and stresses predicted with the ensembles of the 14-th level MTPs and TFMTPs on the training sets created using active learning of MTPs and MatterSim. The results are given with 2-$\sigma$ confidence interval.}
\label{tab:accuracy_on_MatterSim}
\centering
\begin{tabular}{ccccc}
\hline \hline
Model & loss & energy error & force error & stress error \\
& & meV/atom & meV/\AA & eV \\
\hline 
TFMTP & 2.01 $\pm$ 0.15 & $0.63 \pm 0.04$ & $26.8 \pm 0.8$ & $0.232 \pm 0.008$ \\

MTP & 1.78 $\pm$ 0.14 & $0.58 \pm 0.06$ & $25.2 \pm 0.9$ & $0.227 \pm 0.008$ \\
\hline \hline
\end{tabular}
\end{table}

\subsubsection{Training set size and accuracy on DFT data}

At the next stage of our investigation, we actively trained MTP and TFMTP during MD simulations of the FLiNaK system in the NPT ensemble at $T=1200$ K and used DFT calculations (the Quantum Espresso package) for the construction of training sets. Here we actively trained only one potential of each type because the previous test demonstrated a small standard deviation in the results. The number of parameters and configurations for MTP and TFMTP are shown in Table \ref{tab:dataset_size_on_DFT}. We also provide CPU-time in CPU-hours needed to create these training sets. Similarly to Table \ref{tab:dataset_size_on_MatterSim}, TFMTP has approximately 1.5 times fewer parameters, but requires more than two times less amount of configurations for its fitting than MTP.

\begin{table}[!ht]
\caption{Numbers of parameters and configurations in the training sets for the ensembles of the 16-th level MTP and TFMTP actively trained on the data calculated with DFT. CPU-hours spent on the calculation of these training sets. TFMTP has 1.5 times less parameters and its training set size is more than two times fewer than MTP.}
\label{tab:dataset_size_on_DFT}
\centering
\begin{tabular}{cccc}
\hline \hline
Model & \# parameters & \# selected configurations & CPU-hours \\
\hline 
TFMTP & 400 & 844 & 1125 \\

MTP & 608 & 1735 & 2313 \\
\hline \hline
\end{tabular}
\end{table}

Next, we compared the value of the loss function and fitting errors for MTP and similar values estimated with TFMTP on the training set generated with active learning of MTP (see Table \ref{tab:accuracy_on_DFT}). As in the previous case, the errors obtained with TFMTP and MTP are almost indistinguishable and, therefore, TFMTP was effectively trained on the DFT data, without loss of MTP accuracy.

\begin{table}[!ht]
\caption{Values of the loss function and root-mean-square errors for energies, forces, and stresses predicted with the actively trained MTP and TFMTP of the 16-th level on the training set created using active learning of MTP and DFT. The results obtained with TFMTP and MTP are reasonable and in agreement with each other.}
\label{tab:accuracy_on_DFT}
\centering
\begin{tabular}{ccccc}
\hline \hline
Model & loss & energy error & force error & stress error \\
& & meV/atom & meV/\AA & eV \\
\hline 
TFMTP & 0.321 & 1.04 & 32.5 & 0.324 \\

MTP & 0.292 & 0.97 & 31.0 & 0.306 \\
\hline \hline
\end{tabular}
\end{table}

Finally, we tested the transferability of the actively trained MTP and TFMTP models to a higher temperature of 1500 K. To that end, we conducted the DFT-based molecular dynamics simulations in the NPT-ensemble at $T = 1500$ K and $p=1$ bar during 4 ps with a time step of 1 fs. We then took 200 configurations, i.e., each 20-th from this pool of configurations and created a test set. After that we estimated extrapolation grades of these configurations. Both MTP and TFMTP gave grades less than 2, i.e., less than the lower bound $\gamma_{\rm low}$ of permissible extrapolation. In Table \ref{tab:test_errors_on_DFT} we provide test errors on this dataset from which we see that they are close to the ones shown in Table \ref{tab:accuracy_on_DFT}. Thus, TFMTP and MTP fitted on the data for the liquid FLiNaK system demonstrate good transferability to a higher temperature.

\begin{table}[!ht]
\caption{Test energy, force, and stress root-mean-square errors and the loss functions predicted on the test set with the actively trained MTP and TFMTP. The errors and the values of the loss function and close to each other.}
\label{tab:test_errors_on_DFT}
\centering
\begin{tabular}{ccccc}
\hline \hline
Model & loss & energy error & force error & stress error \\
& & meV/atom & meV/\AA & eV \\
\hline 
TFMTP & 0.312 & 1.14 & 29.6 & 0.309 \\

MTP & 0.298 & 1.11 & 29.0 & 0.293 \\
\hline \hline
\end{tabular}
\end{table}

\subsubsection{Values of the loss function, density, and viscosity for ensembles of MLIPs and a comparison with MACE}

For further study, we fitted four more MTPs and TFMTPs on the training sets of 1735 and 844 configurations, respectively, and compared values of the loss function, density, and viscosity calculated with the ensembles of models. The average values of the loss function calculated on the training set obtained with the active learning of MTP for the ensembles of MTPs and TFMTPs are $0.299 \pm 0.006$ and $0.328 \pm 0.005$, and the results are similar to the ones demonstrated above: TFMTP and MTP demonstrate practically identical accuracy. We then compared density and viscosity of FLiNaK computed with these models at $T=600-1200$ K. We also provide the experimental data and the density obtained with MACE-MP-0 \cite{batatia2025_mace_mp_0}.

\begin{figure}[!ht]
	\centering
    \includegraphics[width=1\linewidth]{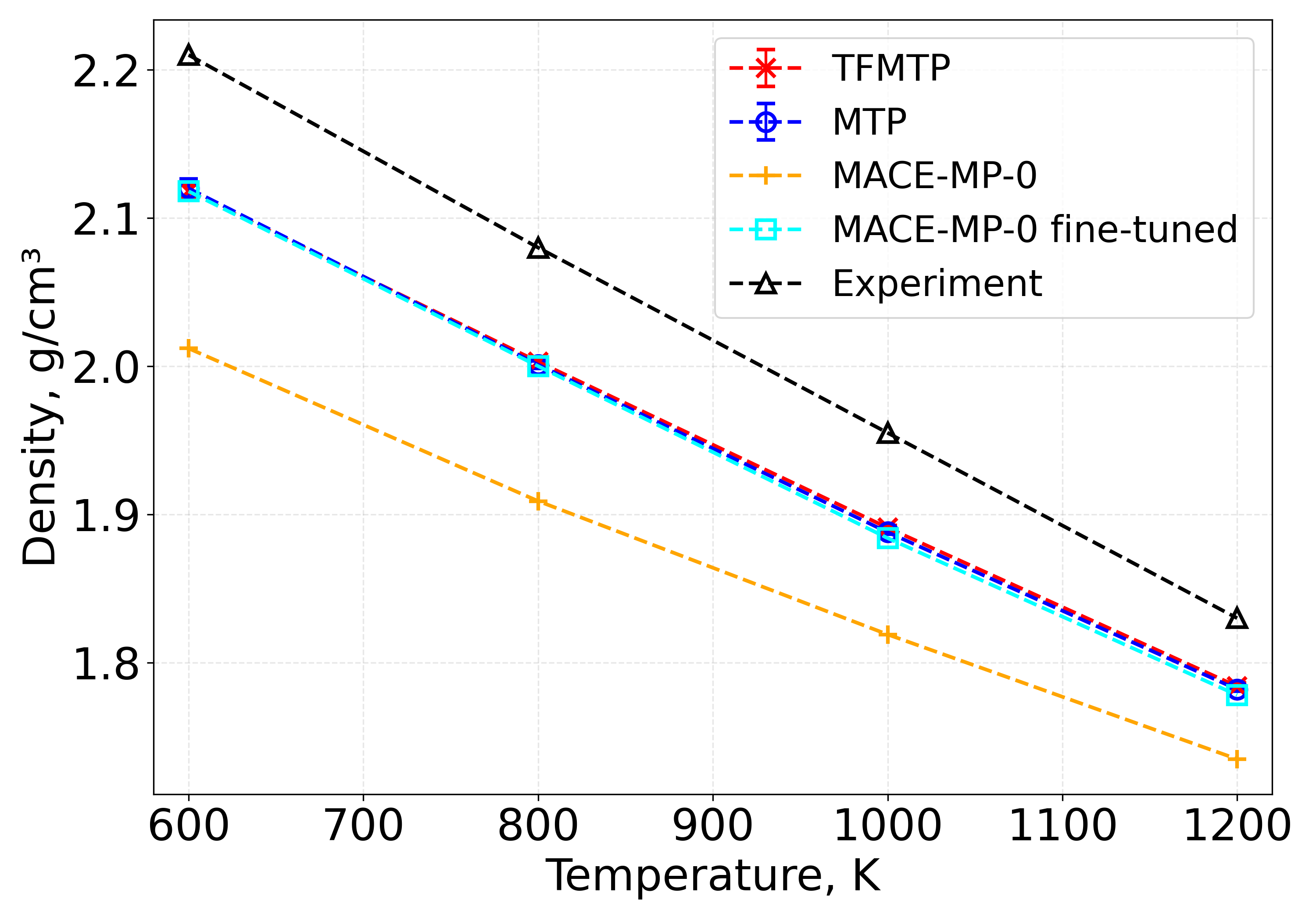}
    \caption{Dependence of the FLiNaK density on temperature obtained with the ensembles of MTP and TFMTP, the MACE-MP-0 models, and experiment (see Ref. \cite{romatoski2017fluoride}). Density calculated with MTP and TFMTP is shown within 2-$\sigma$ confidence interval.}
    \label{fig:density}
\end{figure}

The dependence of the FLiNaK density on temperature, obtained with MTP, TFMTP, the broadly pretrained MACE-MP-0, the fine-tuned MACE-MP-0, and experiment \cite{romatoski2017fluoride} is demonstrated in Fig. \ref{fig:density}. We see identical densities predicted with MTP, TFMTP, and the MACE-MP-0 model fine-tuned on the training set created during the active learning of TFMTP. The deviation between densities calculated with these models and the experimental ones increases from 2.5 \% to 4.5 \% with the decrease of temperature, i.e., it is less than 5 \%, which implies good agreement between MTP, TFMTP, the fine-tuned MACE-MP-0, and the experiment. We note that the deviation between the broadly pretrained MACE-MP-0 and experiment changes from 5 \% to 9 \%, which is about two times greater than that between its fine-tuned version and experiment. This result demonstrates the necessity of fine-tuning foundation models to improve their predictive power.

\begin{figure}[!ht]
	\centering
    \includegraphics[width=1\linewidth]{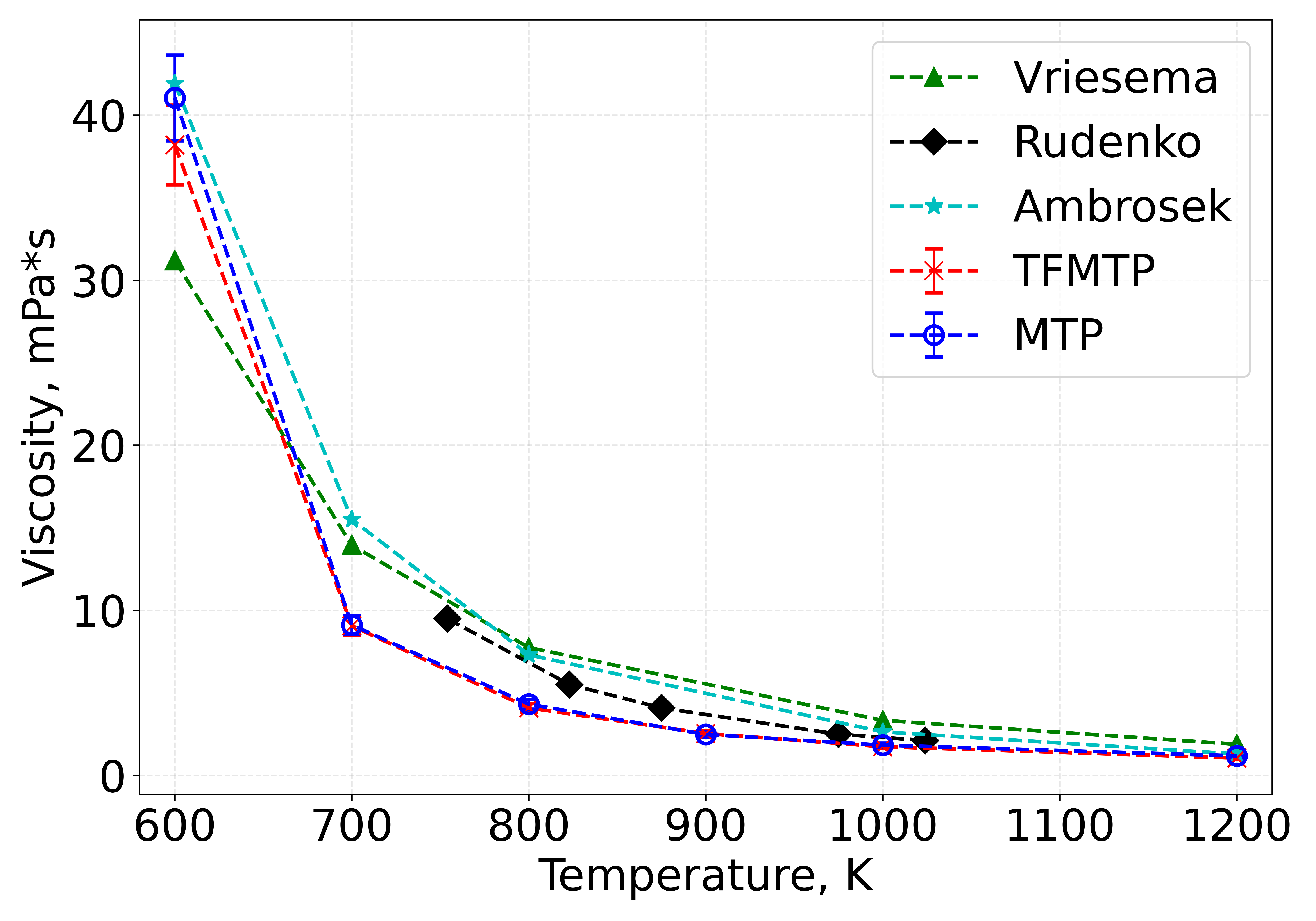}
    \caption{Dependence of the FLiNaK viscosity on temperature obtained with the ensembles of MTP, TFMTP, and experiments (see Ref. \cite{vriesema1979_viscosity,ambrosek2009_visc,rudenko2022_visc}). Viscosity calculated with MLIPs is shown within 2-$\sigma$ confidence interval.}
    \label{fig:viscosity}
\end{figure}

Viscosity calculated with TFMTP and MTP at different temperatures is shown in Fig. \ref{fig:viscosity}. We also provide the experimental results \cite{vriesema1979_viscosity,ambrosek2009_visc,rudenko2022_visc}. As for the previous quantities and properties, viscosities predicted with MTP and TFMTP are indistinguishable. Moreover, they are in qualitative agreement with the viscosities obtained experimentally. We emphasize that we did not calculate viscosity with MACE-MP-0 as the dependencies of density on temperature predicted with MTP, TFMTP, and the fine-tuned MACE-MP-0 are close to each other and because MACE-MP-0 is much more computationally demanding than MTP/TFMTP (see the subsection below).

\subsubsection{Performance of MTP, TFMTP, and MACE}

Besides the properties predicted with MTP, TFMTP, and MACE, we also compared their performance. For this purpose, we estimated CPU-time of molecular dynamics simulations of 120 ps conducted with MTP and TFMTP in the NPT-ensemble at $T=1000$ K for 1512 atoms. We also measured GPU-time needed to conduct exactly the same simulation with MACE-MP-0. Simulations with MTP and TFMTP were conducted on 1 CPU-core of AMD Ryzen Threadripper 7980X, and modeling with MACE-MP-0 was carried out on NVIDIA Tesla V100. The results are given in Table \ref{tab:performance} and they demonstrate the main advantage of the MTP/TFMTP polynomial models over the foundation neural network-based MACE model: only 1 CPU-core is sufficient to conduct 120 ps of molecular dynamics simulation in less than half an hour whereas we need an NVIDIA Tesla V100 and five hours to carry out the same simulation with MACE-MP-0. Thus, MACE-MP-0 is much more computationally demanding than MTP or TFMTP.

\begin{table}[!ht]
\caption{Performance of the 16-th level MTP and TFMTP models estimated in CPU-hours and MACE-MP-0 estimated in GPU-hours. Molecular dynamics simulations of 120 ps in the NPT-ensemble at $T=1000$ K were conducted with the mentioned MLIPs.}
\label{tab:performance}
\centering
\begin{tabular}{ccc}
\hline \hline
Model & Time & Equipment  \\
\hline 
TFMTP & 0.40 CPU-hours & AMD Ryzen Threadripper 7980X \\

MTP & 0.32 CPU-hours & AMD Ryzen Threadripper 7980X \\

MACE & 5.3 GPU-hours & NVIDIA Tesla V100 \\
\hline \hline
\end{tabular}
\end{table}

We also note that TFMTP is about 25 \% slower than MTP due to the current implementation. However, in practice, once the fitting of TFMTP is finished, it can be converted to the MTP form, since the core tensors of its parameters $G^{(k)}$, $k=1,\ldots,4$ can be contracted to recover the standard radial MTP parameters $c^{(\beta)}_{z_i,z_j,\mu}$ using \eqref{eq:TT}. Thus, in this converted form, TFMTP runs as fast as the original MTP.

\subsection{Results for Mo-Nb-Ta-W-V}

In this test, we actively trained MTP and TFMTP during the relaxation of the Mo-Nb-Ta-W-V crystal. We started with 110 Mo$_{11}$Nb$_{11}$Ta$_{11}$W$_{11}$V$_{10}$ structures of 54 atoms each, all sharing the same composition (11 Mo, 11 Nb, 11 Ta, 11 W, and 10 V atoms). They were generated from 10 random atomic arrangements, each yielding one relaxed and 10 perturbed structures. Here, we consider only one model of each type, because for FLiNaK, the deviation in the results from the average is negligible for the ensemble of the models. The resulting training set sizes and the number of parameters for the actively trained MTP and TFMTP models are shown in Table \ref{tab:dataset_size_on_equiatomic_MoNbTaWV}, and the values of the loss function and energy, force, and stress errors estimated on the MTP's training set are given in Table \ref{tab:accuracy_on_equiatomic_MoNbTaWV}. The outcomes closely mirror those observed for the FLiNaK system. Specifically, TFMTP uses 1.5 times fewer parameters than MTP, demands a training set roughly half the size, and yet achieves accuracy that is nearly indistinguishable from that of MTP.

\begin{table}[!ht]
\caption{Number of parameters and training set sizes for the 20-th level MTP and TFMTP actively trained during the relaxation of near-equiatomic Mo$_{11}$Nb$_{11}$Ta$_{11}$W$_{11}$V$_{10}$ structures. TFMTP has 1.5 times less parameters and requires approximately two times fewer configurations for its fitting than MTP.}
\label{tab:dataset_size_on_equiatomic_MoNbTaWV}
\centering
\begin{tabular}{ccc}
\hline \hline
Model & \# parameters & \# selected configurations \\
\hline 
TFMTP & 858 & 696 \\

MTP & 1293 & 1541 \\
\hline \hline
\end{tabular}
\end{table}

\begin{table}[!ht]
\caption{Values of the loss function and energy, force, and stress root-mean-square errors predicted with the actively trained MTP and TFMTP of the 20-th level on the MTP's training set created for the near-equiatomic Mo$_{11}$Nb$_{11}$Ta$_{11}$W$_{11}$V$_{10}$. The demonstrated values of the loss function and errors obtained with TFMTP and MTP are in excellent correspondence.}
\label{tab:accuracy_on_equiatomic_MoNbTaWV}
\centering
\begin{tabular}{ccccc}
\hline \hline
Model & loss & energy error & force error & stress error \\
& & meV/atom & meV/\AA & eV \\
\hline
TFMTP & 0.198 & 0.214 & 27.1 & 0.248 \\

MTP & 0.177 & 0.182 & 25.5 & 0.246 \\
\hline \hline
\end{tabular}
\end{table}

Next, we compared the elastic constants and bulk moduli obtained with MTP, TFMTP, and DFT by averaging over configurations including 54 atoms with 10 random arrangements from the training set. The mentioned mechanical properties are shown in Table \ref{tab:mechanical_properties_54_atoms}. From these data, we conclude that the elastic constants and bulk moduli calculated using MTP and TFMTP are in an excellent agreement with each other and with DFT.

\begin{table}[!ht]
\caption{Elastic constants and bulk moduli of the near-equiatomic Mo$_{11}$Nb$_{11}$Ta$_{11}$W$_{11}$V$_{10}$ crystal calculated with TFMTP, MTP, and DFT. All physical quantities are the averages of 10 randomly arranged and relaxed 54-atom systems. The results are given within 1-$\sigma$ confidence interval.}
\label{tab:mechanical_properties_54_atoms}
\centering
\begin{tabular}{cccccc}
\hline \hline
Model & C$_{11}$, GPa & C$_{12}$, GPa & C$_{44}$, GPa & Bulk modulus, GPa \\
\hline 
TFMTP & 331.6 $\pm$ 1.1 & 160.6 $\pm$ 1.0 & 51.2 $\pm$ 0.6 & 217.6 $\pm$ 1.0 \\

MTP & 331.1 $\pm$ 1.8 & 156.8 $\pm$ 0.8 & 49.4 $\pm$ 0.8 & 214.9 $\pm$ 0.6 \\

DFT & 334.3 $\pm$ 5.1 & 159.8 $\pm$ 2.7 & 56.6 $\pm$ 4.5 & 217.9 $\pm$ 1.3 \\
\hline \hline
\end{tabular}
\end{table}

Once we found that the TFMTP and MTP models enable predicting mechanical properties of the near-equiatomic Mo$_{11}$Nb$_{11}$Ta$_{11}$W$_{11}$V$_{10}$, we investigated their transferability to the unseen compositions described in Computational details. First, we estimated the extrapolation grades of 80 configurations prerelaxed with DFT, including eight compositions and 10 arrangements of atoms for each composition. We observed that all the extrapolation grades exceed 10000 for both MTP and TFMTP. Therefore, using the training sets from Table \ref{tab:dataset_size_on_equiatomic_MoNbTaWV} as a starting point, we actively trained TFMTP and MTP during the relaxation of 80 configurations with eight different compositions and found that TFMTP selected 160 additional configurations (856 configurations in total), while MTP selected 293 additional configurations (1834 structures in total). Thus, the number of selected configurations is proportional to the number of parameters in the models. Furthermore, the accuracy of MTP and TFMTP remains comparable (see Table \ref{tab:accuracy_on_MoNbTaWV}). We also emphasize that here we spent about 17120 CPU hours creating a training set for TFMTP and about 36680 CPU hours generating a training set for MTP. Generating these datasets is more computationally expensive than for FLiNaK because the k-mesh size here is $4 \times 4 \times 4$ instead of $2 \times 2 \times 2$ as in the case of FLiNaK. Therefore, active learning of the compressed models is particularly important for metallic materials, or for systems where more computationally expensive functionals than PBE are required.

\begin{table}[!ht]
\caption{Values of the loss function and energy, force, and stress root-mean-square errors predicted with the actively trained MTP and TFMTP of the 20-th level on the MTP's training set created for different Mo-Nb-Ta-W-V compositions. The demonstrated values calculated with TFMTP and MTP are close to each other.}
\label{tab:accuracy_on_MoNbTaWV}
\centering
\begin{tabular}{ccccc}
\hline \hline
Model & loss & energy error & force error & stress error \\
& & meV/atom & meV/\AA & eV \\
\hline
TFMTP & 0.254 & 0.413 & 27.8 & 0.294 \\

MTP & 0.208 & 0.320 & 25.5 & 0.259 \\
\hline \hline
\end{tabular}
\end{table}

\begin{figure}[!ht]
	\centering
    \includegraphics[width=1\linewidth]{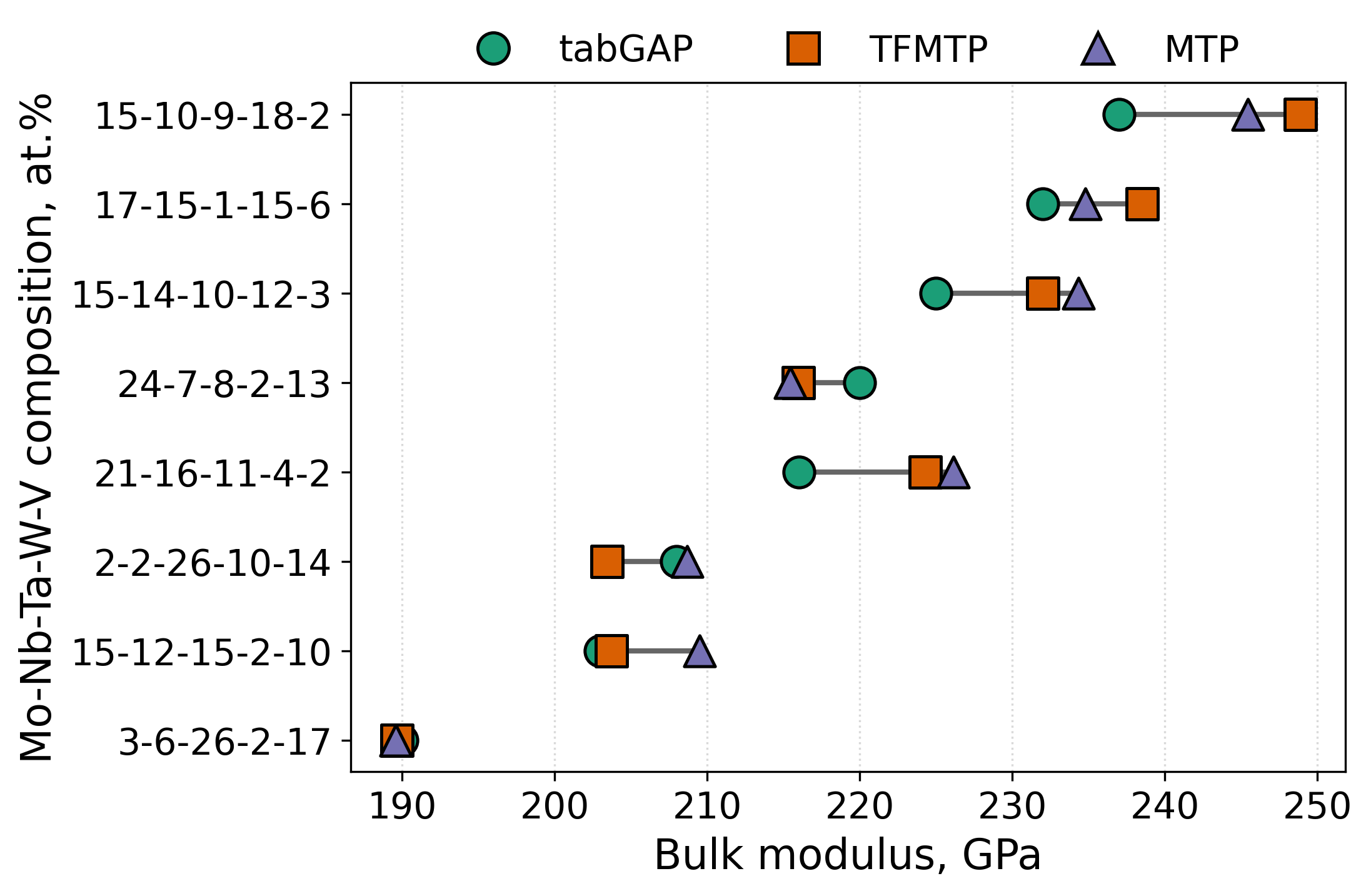}
    \caption{Bulk moduli calculated with TFMTP, MTP, and tabGAP for eight compositions of Mo-Nb-Ta-W-V.}
    \label{fig:bulk_moduli}
\end{figure}

Finally, we calculated the bulk moduli of the eight new compositions with MTP and TFMTP and compared them with the ones obtained using the tabGAP model \cite{byggmastar2021_monbtawv}. The results are shown in Fig. \ref{fig:bulk_moduli}. TFMTP and MTP yielded close bulk modulus values, with a maximum difference of about 6 GPa. These values are also in rather good agreement with the tabGAP model: the largest discrepancy is approximately 12 GPa, which amounts to a relative error of only about 5 \% with respect to the tabGAP values. Notably, all three models gave an overall similar ranking of compositions based on bulk moduli, with the only pronounced difference observed for Mo$_{24}$Nb$_{7}$Ta$_{8}$W$_{2}$V$_{13}$ and Mo$_{21}$Nb$_{16}$Ta$_{11}$W$_{4}$V$_{2}$: the tabGAP model predicts a larger bulk modulus for Mo$_{24}$Nb$_{7}$Ta$_{8}$W$_{2}$V$_{13}$, whereas MTP and TFMTP predict a bulk modulus 11 and 8 GPa larger, respectively, for Mo$_{21}$Nb$_{16}$Ta$_{11}$W$_{4}$V$_{2}$. Overall, the bulk moduli computed with TFMTP and MTP after active training on the previously unseen Mo-Nb-Ta-W-V compositions are in good agreement with the tabGAP predictions, demonstrating that the active learning procedure enables reliable transferability of both models to new compositions.

\subsection{Discussion}

The aim of this work was to verify whether the compressed TFMTP model requires less DFT data for its fitting than the original MTP model while maintaining the same accuracy. For the four-component FLiNaK system and the five-component Mo-Nb-Ta-W-V random alloy, we demonstrated that TFMTP which has 1.5 times fewer parameters, requires two times less amount of configurations for its fitting, and preserves the accuracy of MTP. A drawback of the current implementation of the iterations of the active learning algorithm is the problem with TFMTP scaling: we have to choose an interval for the initial guess for proper TFMTP fitting, as for some intervals, the BFGS algorithm gets stuck at the line search step. Due to this reason, we chose smaller weights in the loss function (see the Computational details section) than the classical $w_{\rm e}=1$ eV$^{-2}$, $w_{\rm f}=0.01$ (eV / \AA)$^{-2}$, and $w_{\rm s}=0.001$ eV$^{-2}$ that are typically chosen for MTP fitting and subtracted the minimum energy of configurations in the initial training sets to reduce the value of the initial loss function for the FLiNaK system. For Mo-Nb-Ta-W-V, we subtracted the per-atom energy from the total energy of each configuration and fit to this subtracted energy, which makes the target quantity of energy almost independent of the composition. Such a technique enables fitting TFMTP starting from an initial guess from the interval (0.1, 0.8) for FLiNaK and from the interval (0.05,0.4) for Mo-Nb-Ta-W-V with a uniform distribution at each iteration of the active learning algorithm. In our further studies, we plan to implement an automated algorithm for TFMTP scaling, but this is beyond the scope of this work.

Although this work focuses on MTP, parameter compression by tensor factorization is, in principle, applicable to other MLIPs when suitable parameter tensors admit low-rank representations that preserve the model's symmetries and predictive accuracy. Related constructions already appear in neural network potentials. Darby et al.~\cite{darby2023_tensor_reduced} developed tensor-reduced atomic density representations and described their use in MACE~\cite{batatia2022_mace}. Equivariant tensor network potentials also incorporate tensor factorizations into their parameterization~\cite{hodapp2024-ETN}.

The present study examines the effect of compression on training sets generated by the MaxVol-based active learning for MTP. In the formulation used here, the submatrix of the maximum volume contains derivatives with respect to all model parameters, including the radial parameters, which constitute about 80 \% of the MTP parameters for the models considered here. Therefore, compressing the radial parameters reduces the number of columns in this matrix. For the systems considered, our experiments show that TFMTP is trained on fewer selected configurations than MTP while achieving comparable accuracy. Active learning for neural network potentials uses different approaches, including ensemble disagreement in DP-GEN for DeePMD~\cite{wang2018-deepmd,zhang2020_dpgen} and criteria based on last-layer gradients or random projections of gradient features~\cite{zaverkin2022_batch_AL}. Assessing how tensor compression affects training set size in these settings involves both the network architecture and its interaction with uncertainty estimation and configuration selection. This question lies beyond the scope of the present work, which focuses on MTP within the MaxVol-based active learning framework.

\section{Conclusions}

In this study, we demonstrated that the low-rank approximation of the Moment Tensor Potential (MTP), namely, the tensor-factorized MTP (TFMTP), requires fewer configurations for fitting than the original MTP while preserving its accuracy. To that end, we explicitly implemented the functional form of TFMTP in the program code. Next, we actively trained both MTP and TFMTP using the MaxVol-based algorithm proposed in Ref. \cite{podryabinkin2017_AL} during atomistic simulations of two benchmark systems: a four-component molten salt mixture, LiF-NaF-KF (FLiNaK), and a five-component Mo-Nb-Ta-W-V random alloy. We then compared the training set sizes required to fit the models, assessed the accuracy of the trained models on the training sets created with active learning of MTP, and evaluated the density and viscosity of the FLiNaK system at $T=600-1200$ K, alongside the elastic properties of the Mo-Nb-Ta-W-V crystal at zero temperature.

For the FLiNaK system, we initiated our investigation by fitting the potentials to configurations labeled using the MatterSim model \cite{yang2024mattersim}. We found that an ensemble of actively trained, 1.5-fold compressed TFMTPs of the 14th level required over 1.5 times fewer configurations for training than the original MTP, while maintaining closely matching values of the loss function and energy, force, and stress errors. After that, we fitted the ensembles of MTPs and 1.5-fold compressed TFMTPs of the 16th level on density functional theory (DFT) data. We demonstrated that the values of the loss function, density, and viscosity obtained with MTPs and TFMTPs were in excellent agreement with each other, and that the actively trained TFMTP yielded more than a two-fold reduction in the number of configurations compared to MTP. Furthermore, the calculated temperature dependencies of density and viscosity are in perfect agreement with the most recent experimental works~\cite{romatoski2017fluoride, rudenko2022_visc} and in reasonable agreement with the previously obtained literature results. We also compared the density calculated using MTP/TFMTP with that calculated with the broadly pretrained and the fine-tuned MACE-MP-0 model \cite{batatia2025_mace_mp_0}. The density predicted with the MACE-MP-0 model fine-tuned on the training set obtained during active learning of TFMTP is in excellent agreement with TFMTP/MTP and experiment, but the TFMTP and MTP models required significantly less CPU-time than MACE did on GPU for the density calculation. The values of the density computed with the broadly pretrained MACE-MP-0 model are further from the experimental data which demonstrates that creating a training set for fine-tuning of the MACE foundation model is necessary. Finally, we demonstrated the transferability of the TFMTP and MTP trained at $T=1200$ K to a higher temperature of 1500 K. 

In the case of the Mo-Nb-Ta-W-V crystal, we started our investigation from the near-equiatomic Mo$_{11}$Nb$_{11}$Ta$_{11}$W$_{11}$V$_{10}$ crystal and fitted the 20th-level MTP and TFMTP only on the DFT data, and chose elastic constants and bulk moduli as the target properties. We obtained results consistent with the FLiNaK system: TFMTP has 1.5 times fewer parameters and requires an approximately two times smaller training set size, while its values of the loss function, energy, force, stress errors, elastic constants, and bulk modulus remain indistinguishable from the MTP predictions. Additionally, we calculated the elastic constants and bulk modulus using direct DFT simulations and observed a good correspondence with those obtained via the fitted potentials. Next, we investigated the transferability of the TFMTP and MTP models to the eight unseen compositions of Mo-Nb-Ta-W-V that were not included in the training sets. We found that both TFMTP and MTP extrapolate for all these unseen compositions. Therefore, we actively trained TFMTP and MTP during the relaxation of structures with these new compositions. The number of configurations in the training sets, which initially included only one composition, increased proportionally to the number of parameters of the potential, and the ratio of the number of configurations in the resulting TFMTP and MTP training sets remained close to two. Finally, we compared the bulk moduli calculated with MTP, TFMTP, and the tabGAP model \cite{byggmastar2021_monbtawv}. The bulk moduli obtained with TFMTP and MTP are close to each other and overall are in good agreement with the tabGAP predictions. Thus, the active learning procedure for both TFMTP and MTP yields reliable transferability to new compositions.

In future work, we plan to develop a method for the optimization of TFMTP with automated scaling. This algorithm will enable us fitting TFMTP without the need to carefully choose an initial guess or to scale the loss function.

\begin{acknowledgments}
The work was supported by the grant for research centers in the field of AI provided by the Ministry of Economic Development of the Russian Federation in accordance with the agreement 000000C313925P4E0002 and the agreement with HSE University No 139-15-2025-009. This research was supported in part by computational resources of HPC facilities at the HSE University~\cite{kostenetskiy2021hpc}.

The authors acknowledge Dmitry Korogod for fruitful discussions regarding this study and Alexey Kotykhov for the calculations conducted with the MACE model.
\end{acknowledgments}

\section*{Author declarations}

\subsection*{Conflict of interest}

The authors have no conflicts to disclose.

\subsection*{Author Contributions}

\textbf{Anna Bondarenko}: Data curation (supporting); Formal analysis (supporting); Software (lead). \textbf{Nikita Rybin}: Formal analysis (supporting); Methodology (supporting); Visualization (equal); Writing - review \& editing (supporting). \textbf{Maxim Rakhuba}: Conceptualization (equal); Formal analysis (supporting); Methodology (equal); Supervision (supporting); Writing - review \& editing (supporting). \textbf{Ivan S. Novikov}: Data curation (lead); Visualization (equal); Conceptualization (equal); Formal analysis (lead); Methodology (equal); Software (supporting); Supervision (lead); Writing - original draft (lead); Writing - review \& editing (lead).

\section*{Data Availability Statement}

Data will be made available on request.

\end{document}